\documentstyle[twocolumn,prl,aps,epsf,epsfig]{revtex}
\input epsf

\begin{document}
\draft
\twocolumn[\hsize\textwidth\columnwidth\hsize\csname@twocolumnfalse\endcsname 

\title{Theory of Nonlinear Susceptibility and Correlation Length in
Glasses and Liquids} \author{  Claudio
Donati$^{1,2}$, Silvio Franz$^3$, Sharon C. Glotzer$^1$ and Giorgio Parisi$^2$}
\address{  $^1$Center for Theoretical and Computational
Material Science, and Polymers Division, National Institute of
Standards and Technology, Gaithersburg, Maryland 20899 USA\\
$^2$Universit\`a di Roma ``La Sapienza'' P.le A. Moro 2, 00185 Rome,
Italy, \\
$^3$The Abdus Salam ICTP, Strada Costiera 11, P.O. Box 563,
34100 Trieste, Italy\\}

\date{Revised version resubmitted to Physical Review Letters: \today}

\maketitle
\begin{abstract}
Within the framework of the effective potential theory of the
structural glass transition, we calculate for the $p$-spin model and a
hard sphere liquid in the hypernetted chain approximation a static
nonlinear susceptibility related to a four-point density correlation
function, and show that it grows and diverges in mean field with
exponent $\gamma=1/2$ as the critical temperature $T_c$ is approached
from below.  When $T_c$ is approached from above, we calculate for the
$p$-spin model a time dependent nonlinear susceptibility and show that
there is a characteristic time where this susceptibility has a
maximum, and that this time grows with decreasing $T$.  We find that
this susceptibility diverges as $T_c$ is approached from above, and
has key features in common with the ``displacement-displacement
susceptibility'' recently introduced to measure correlated particle
motion in simulations of glass-forming liquids.
\end{abstract} 

\vskip2pc]
\narrowtext

Tempted by the possibility of treating the glass transition within the
framework of conventional critical phenomena, researchers have long
searched for evidence of a static correlation length that becomes
large as the glass transition is approached. However, no clear
evidence for such a length has been reported \cite{leheny}. Recent
numerical studies of glass-forming liquids have identified a dynamical
length associated with the range over which particle motions are
correlated \cite{hiw,heuer,dgp,bdbg,kdppg,dgpkp}.  By introducing a
``displacement-displacement'' correlation function and generalized
``displacement susceptibility'' $\chi_u(t)$, Refs.~\cite{dgp,bdbg}
showed in two different model liquids that this length grows with
decreasing temperature $T$ as the mode coupling temperature $T_c$ is
approached from above, despite the fact that density and composition
correlations remain short-ranged.  Evidence for a growing length
associated with correlated particle motion at fixed $T$ in simulations
below $T_c$ has also been reported \cite{PARI}.

In this Letter we calculate within the effective potential theory a new 
susceptibility $\chi_4$ associated with a four-point density
correlation function \cite{dasgupta91}. We show that below $T_c$,
$\chi_4$ diverges with exponent $1/2$ as $T \to T_c^-$.  We show that
above $T_c$, $\chi_4$ is time-dependent and resembles the
susceptibility calculated in \cite{dgp,bdbg}.  In particular, we show
that $\chi_4(t)$ has a maximum which diverges in mean field as $T \to
T_c^+$. We argue that (1) the diverging correlation length implied by
the diverging susceptibility is associated with incipient ergodicity
breaking at $T_c$, and (2) this length underlies the growing range of
correlated particle displacements measured in Refs.~\cite{dgp,bdbg}.
Finally, we test our theoretical predictions above $T_c$ using data
from molecular dynamics simulations of a model glass-forming liquid.

We use two different approaches depending on $T$. In the low $T$
regime ($T<T_c$), we calculate a static, nonlinear susceptibility
using the effective potential theory \cite{pot}, and in the high $T$
regime ($T>T_c$), we calculate a dynamic, nonlinear susceptibility in
a dynamical approach.  The low $T$ calculations are performed both for
a hard-sphere liquid in the hypernetted chain (HNC) approximation
\cite{hnc} and for the spherical $p$-spin model \cite{pspin}; the
dynamical calculations are performed only for the spherical $p$-spin
model.  This is the simplest model that (i) allows both static and
dynamic quantities to be calculated exactly, and (ii) exhibits several
key features common to liquids in and close to their glassy regime
\cite{hnc,CFP,mp-last}. For example, the high-$T$ dynamics of the
$p$-spin model is described exactly by the ideal mode coupling
equations \cite{kirtir,horner} and displays a sharp transition at
$T_c$. 

In the effective potential theory, below $T_c$ phase space is split
into separate ergodic components which remain stable at all
temperatures (this stability is due to the mean field nature of the
theory). A key hypothesis of our approach is that in real systems the
ergodic components are {\it metastable} states with a long but finite
escape time. In this framework we can use a static approach to compute
the properties of these metastable states below $T_c$. We can also
obtain information on the dynamical properties above $T_c$, where mean
field theory predicts that the time to escape from a metastable state
diverges on approaching $T_c$. The divergence of a {\it static}
susceptibility inside the metastable state when we approach $T_c$ from
below is directly related to the divergence of the corresponding
{\it dynamical} susceptibility when we approach $T_c$ from above. As is
typical for mean field theories, which neglect spatial fluctuations, a
diverging correlation length is deduced from the diverging
susceptibility.

We first describe the essential elements of the effective potential
theory (a complete description can be found in \cite{CFP}).  The
theory is formulated using a measure of the similarity or ``overlap''
$q$ between two configurations $X$ and $Y$ as an order parameter to
detect vitrification.  Different definitions of $q$ can be used in
different systems and the main results of the theory do not depend on
the particular definition adopted. In the case of simple liquids with
$N$ particles at fixed density \cite{CFP,P}, one can define
\begin{eqnarray}
q(X,Y)=&&\frac{1}{N} \sum_{{i=1}\atop{j=1}}^{N} w(x_i-y_j)
\label{Q}\\
=&&{1\over N} \int dx\, dy \, w(x-y) {\rho}{_X(x)} {\rho}{_Y(y)},\nonumber
\end{eqnarray}
where $X=\{x_1,...,x_N\}$, $Y=\{y_1,...,y_N\}$, and
${\rho}{_Z(x)}=\sum_{i=1}^N \delta (x-z_i)$ is the microscopic density
corresponding to the configuration $Z=X,Y$. Here $w(r)$ is chosen to
be a smooth, continuous, short-range function close to one for $r< a
r_0$ and close to zero otherwise ($r_0$ is the radius of a
particle). The value of $a<1$ is arbitrary, and $a=0.3$ is a good
compromise for an overlap insensitive to small thermal fluctuations
\cite{CFP,P}.

\begin{figure}[tbp]
\hbox to\hsize{\epsfxsize=1.0\hsize\hfil\epsfbox{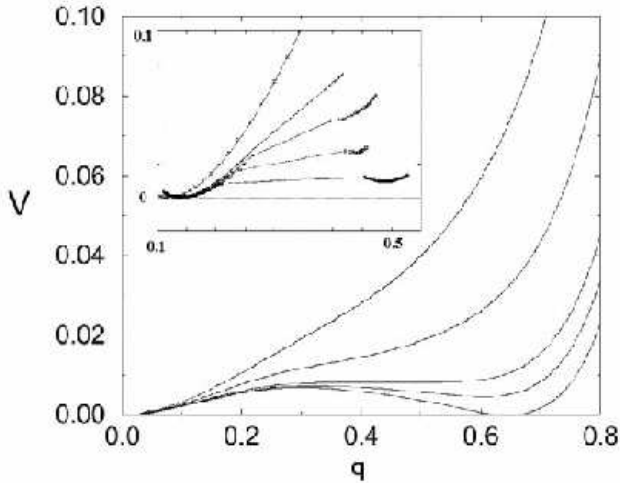}\hfil}
\caption[0]{\protect\label{fig1} The effective potential $V(q)$ for
the $p$-spin model, for several values of $T$. At high $T$
the potential is everywhere convex, and at low $T$, $V(q)$ exhibits
two minima. In the inset we show the effective potential for a hard
sphere fluid in the HNC approximation for several values of the
density ($\rho=1.0,1.14,1.17,1.19,1.20$). Here the potential is
calculated around the high- and in the low-q minima, and the lines
joining the two minima are guides to the eye \cite{CFP}.}
\end{figure} 

The effective potential $V(q)$, which is a constrained free energy, is
constructed by choosing a fixed reference equilibrium configuration
$Y$ at temperature $T$, and calculating the free energy of a
configuration $X$ that has an overlap $q$ with $Y$:
\begin{equation}
V(q)=-\frac{T}{N} \log \int d X \exp(-\beta H(X)) \delta(q(X,Y)-q).
\end{equation}
Here $\beta \equiv 1/k_BT$ and $H(X)$ is the potential energy of $X$.
$V(q)$ is self-averaging with respect to the choice of $Y$.

The typical mean-field shape of $V(q)$ for a system undergoing
vitrification is shown in Fig.~\ref{fig1} for several values of $T$.
The shape of $V(q)$ allows one to distinguish the liquid from the
glassy phase since the presence of a single or multiple minima
indicates either ergodicity or broken ergodicity, respectively.  At
high $T$, the system is ergodic and $V(q)$ is convex, with a single
minimum at a small value of the overlap $q$ between any two
configurations chosen with the Boltzmann weight.  Upon lowering $T$,
the curvature changes sign, and at $T_c$, $V(q)$ develops a secondary
minimum at a higher value of $q$. This signals breaking of ergodicity:
at $T_c$ the configuration space become disconnected into an
exponentially large number of ``ergodic components'' ${\cal N}\sim
\exp(N \Sigma)$, each carrying vanishing weight in the Boltzmann
distribution\cite{CFP}. A fundamental result of the theory, central in
the following discussion, is that physical quantities calculated in
the primary minimum represent averages computed with the Boltzmann
weight, while the same quantities calculated in the secondary minimum
represent averages computed only within a single ergodic component.

To calculate physical quantities in the effective potential theory, 
we introduce the Legendre transform of $V(q)$: $
\Gamma(\epsilon)=\min_q V(q)-\epsilon q$, where $\epsilon$ is a
``field'' conjugate to $q$, and corresponds to a coupling between
configurations.  For example, the average overlap $\langle q\rangle$
can be computed as $ \langle q\rangle =\frac{\partial \Gamma}{\partial
\epsilon}|_{\epsilon\to 0}$, where $\langle \cdots \rangle$ represents
either of the two types of averages. The overlap susceptibility is
defined as
\begin{equation}\label{DQDEPS}
\chi_4 =\frac{\partial \langle q \rangle}{\partial \epsilon}\Big|_{\epsilon\to 0} = \beta N \Big( \langle q^2 \rangle- \langle q \rangle^2 \Big),
\end{equation}
where $q\equiv q(X,Y)$.  Inserting Eq.~\ref{Q} in Eq.~\ref{DQDEPS}
allows us to rewrite
$\chi_4$ as
\begin{eqnarray}
\chi_4= N \beta\int d x_1 d y_1d x_2 d y_2 w(x_1-y_1) w(x_2-y_2)\nonumber\\
G_4(x_1,y_1,x_2,y_2),
\label{chi}
\end{eqnarray}
where we have defined the four-point density correlation function \cite{dasgupta91},
\begin{eqnarray}
G_4(x_1,y_1,x_2,y_2)= &&
{1\over N^2} \Big[\big\langle {\rho}{_X(x_1)}{\rho}{_Y(y_1)}{\rho}{_X(x_2)}
{\rho}{_Y(y_2)}
\big\rangle\nonumber\\
&& -
\big\langle {\rho}{_X(x_1)}{\rho}{_Y(y_1)}\rangle\langle
{\rho}{_X(x_2)}
{\rho}{_Y(y_2)}
\big\rangle\Big].
\label{Gx1y1x2y2}
\end{eqnarray}

The two types of averages for $\chi_4$ are easily calculated.  We find
that when calculated with respect to the Boltzmann average, $\chi_4$
is regular (and small) at all $T$. However, when calculated within the
secondary minimum (i.e. averaged within a single ergodic component)
$\chi_4$ grows for increasing $T$, and diverges at $T_c$ as a power
law $\chi_4\sim (T_c-T)^{-\gamma}$, as shown in Fig.~2.  This
demonstrates that equilibrium configurations within a single ergodic
component are highly correlated, while configurations in different
components are not.  In both the $p$-spin model and hard sphere model
in the HNC approximation, the form of $V(q)$ is cubic around the
second minimum, and thus the value of the exponent $\gamma$ is equal
to $1/2$ (i.e. the coefficient of the quadratic term resulting when
$V(q)$ is expanded around the second minimum vanishes as
$(T_c-T)^{1/2}$). This value is universal within mean field and
coincides with the value of $\gamma$ for mean-field spinodal
transitions.  If we assume usual scaling with this value of $\gamma$
then the correlation length exponent $\nu$ is related to the anomalous
dimension $\eta$ by $\nu=1/2(2-\eta)$.  We emphasize that the mean
field level at which we are describing the system should be the same
as that of the ideal mode coupling theory. The success of MCT in
predicting relations between exponents of various dynamical quantities
\cite{MCT} leads us to speculate that the mean-field value of
$\gamma$, or a close value, could be observed in real systems, and we
test this using MD simulations at $T>T_c$ later in this paper.

\begin{figure}[tbp]
\hbox to\hsize{\epsfxsize=1.0\hsize\hfil\epsfbox{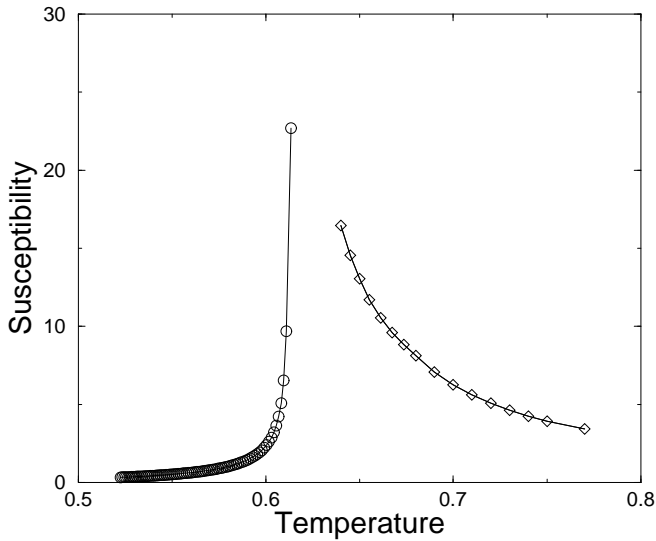}\hfil}
\caption[0]{\protect\label{fig3} The static susceptibility $\chi_4$
(circles) calculated at low temperature ($T<T_c$), plotted vs. $T$,
together with the maximum of the time dependent susceptibility
$\chi_4(t)$ (diamonds) at high temperature ($T>T_c$). The solid line
through the low-$T$ data is included as a guide to the eye. The dashed
line through the high-$T$ data indicates a power law fit:
$\chi_4(T)=a/(T-T_c)^{1/2}+b$ .  In the $p$-spin model, $T_c = 0.612$
\cite{pspin} and the fit gives $a=3.67$, $b=-6.28$.  In real systems
one can expect a rounding of the divergence at $T_c$.}
\end{figure} 

We now turn to the high temperature region $T>T_c$ where the system is
ergodic. In this region there is no secondary minimum in $V(q)$, and
the static susceptibility does not exhibit any singular behavior.
However, in this temperature regime particles of a supercooled liquid
``oscillate'' within cages formed by their neighbors, and the system
is effectively ``frozen'' for a characteristic time which grows and
diverges as $T_c$ is approached. This transient localization
corresponds to highly correlated regions of phase space that have
finite lifetime and represent the high temperature precursors of the
low temperature ergodic components. 

We study the dynamics of such a system starting in an equilibrium
initial condition $Y=X(0)$ and evolving in time with potential energy
${\cal H} = H[X]-\epsilon q(X,Y)$, and we calculate the dynamic
susceptibility $\chi_4(t)$ associated with the time dependent overlap
$q(t)\equiv q(X(t),X(0))$ from Eqs.~4 and 5, with $q\equiv q(t)$.  To
calculate $\chi_4(t)$ above $T_c$ we use the mode coupling approximation
(again the calculations are performed for the spherical $p$-spin
model, where MCT is exact).  This gives rise to closed equations for
the time-dependent overlap correlation and response functions $C(t,t')=\langle
q(X(t),X(t'))\rangle$ and $R(t,t')=\frac {\delta \langle
x_i(t')\rangle }{\delta h_i(t)}$ which are a slight generalization of
the equations discussed in Ref.~\cite{pot} (the details of the
calculation will be given elsewhere). We solve for
$\chi_4(t)=\frac{\delta C(t,0)}{\delta \epsilon}$ by integrating these
equations numerically as in \cite{pot} for different values of $T$.
As shown in Fig.~3a, we find that $\chi_4(t)$ displays a maximum as a
function of time, which increases and shifts to larger $t$ as $T \to
T_c^+$.

\begin{figure}[tbp]
\hbox to\hsize{\epsfxsize=1.0\hsize\hfil\epsfbox{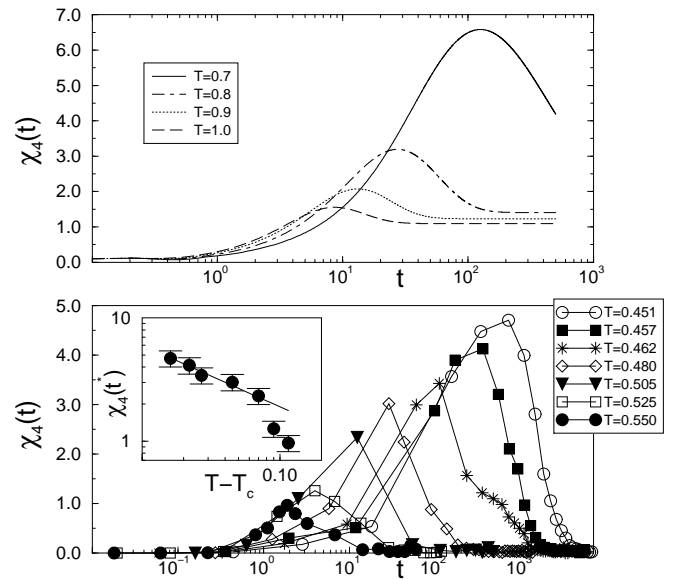}\hfil}
\caption[0]{\protect\label{fig4} (a) The nonlinear susceptibility
$\chi_4(t)$ computed from the theory in the $p$-spin model with $p=3$,
for temperatures $T=0.7,0.8,0.9,1.0$ ($T_c=.612$). The long time limit
corresponds to the static susceptibility which for this model above
$T_c$ is equal to $\chi_4(\infty)=1/kT$.  (b) The nonlinear,
time-dependent susceptibility $\chi_4(t)$ calculated for the LJ binary
mixture above $T_c$.  The long time limit for the liquid is negligable
due to the normalization.  Inset: The maximum $\chi_4(t_4^*)$ plotted
as a function of $T-T_c$, with $T_c=0.435$ \cite{kdppg,dgpkp,kob}.
The solid line indicates a power law fit to $\chi_4(t_4^*) \sim
(T-T_c)^{1/2}$, and is included in order to compare the simulation
data with the analytical mean-field prediction.}
\label{figure3}
\end{figure} 

The temperature dependence of the maximum of $\chi_4(t)$ is shown in
Fig.~2 (circles); we find that the maximum behaves as a negative power
of $T-T_c$ on approaching $T_c$.  Although we did not attempt to
compute the value of $\gamma$ above $T_c$ analytically, in analogy
with spinodal points we expect that the values of the exponents on the
two sides of the transition would be the same. Indeed, we find that
the data are compatible with the mean-field value of $1/2$ calculated
below $T_c$ (a power-law fit to the high-$T$ data with $T_c$ fixed and
$\gamma$ as a free parameter gives $\gamma=0.52$).  Thus we find that
the dynamic theory of the $p$-spin model predicts a diverging dynamic,
nonlinear susceptibility, (and thus by standard arguments a diverging
dynamical correlation length), associated with a four-point density
correlation function.

In real systems, we expect that the transition at $T_c$ will be
smeared by the existence of dynamical processes that restore
ergodicity, and which are not taken into account in the mean field
approach.  This has two important ramifications.  The first is that
even below $T_c$, $\chi_4(t)$ should display a maximum at finite
time. The second is that the divergence of $\chi_4$ as a function of $T$
should be smoothed.

To test our predictions, we calculate $\chi_4(t)$ for the same 80:20
Lennard-Jones liquid (containing 8000 particles) studied in
Ref.~\cite{dgp,kdppg,dgpkp}.  Complete details of the simulation may be
found in \cite{dgpkp}. We evaluate $\chi_4(t)$ using the
time-dependent generalization of Eq.~3, by calculating the
fluctuations in the overlap $q$ measured between two configurations of
the system separated by a time $t$.  In Fig.~3 we show $\chi_4(t)$ as
a function of the time interval $t$ between the two equilibrium
configurations, for seven different values of $T$ approaching
$T_c=0.435$ from above.  In qualitative agreement with our theoretical
predictions, we find that for the binary Lennard-Jones liquid,
$\chi_4(t)$ has a maximum $\chi_4(t_4^*)$ at an intermediate time
$t_4^*$.  The amplitude of the peak grows and shifts to longer times
with decreasing $T$.  As shown in the inset, the $T$-dependence of
$\chi_4(t_4^*)$ is compatible with the mean field prediction. However,
we caution that a rigorous test of the theory would require additional
simulations closer to $T_c$ and improved statistics.

In summary, we have calculated both within the effective potential
theory and in a dynamical approach, a diverging susceptibility below
and above the mode coupling dynamical critical temperature,
respectively.  As seen clearly from Eq.~4, this susceptibility is
related to the growing range of a four-point, time-dependent density
correlation function.  Our findings suggest \cite{gdp,comment} an
interpretation of the physics underlying the displacement-displacement
correlation function calculated in Refs.~\cite{dgp,bdbg} in terms of a
four-point density correlation function.  The correlation function
calculated in Refs.~\cite{dgp,bdbg} measures the extent to which the
(scalar) displacements of a pair of particles separated by a distance
$r$ are spatially correlated. Specifically, this function is similar
to the static, two-point pair correlation function $g(r)$, but with
each particle's contribution to $g(r)$ weighted by its subsequent
displacement over a time interval $[0,t]$.  In contrast, the
four-point function $G_4$ studied in the present work measures the
extent to which ``overlapping'' particles within a time interval
$[0,t]$ are correlated. Although these two correlation functions are
different by definition (and thus, e.g. the exponents describing the
$T$-dependence of the generalized susceptibilities $\chi_4$ and
$\chi_U$ will be different), we believe that the length scale
associated with $G_4$ underlies the growing range of correlated
particle displacements measured in Ref.~\cite{dgp,bdbg}.  A further
critical test of our theoretical predictions would consist in the
simulation or experimental measurement of $\chi_4(t)$ or the related
correlation length below $T_c$.  The quantities studied here should be
measurable experimentally in colloidal liquids using confocal video
microscopy \cite{colloids}.

S.F. would like to acknowledge the kind hospitality of the NIST Center
for Theoretical and Computational Materials Science where part of this
work was accomplished.

\end{document}